4

# Magnetization of Flat Superconducting Films on Ferromagnetic Substrates

Leonid Prigozhin[1] and Vladimir Sokolovsky[2]

[1]Blaustein Institutes for Desert Research, Ben-Gurion University of the Negev, Sde Boqer Campus, 84990 Israel
[2]Physics Department, Ben-Gurion University of the Negev, Beer-Sheva, 84105 Israel

**Ferromagnetic substrate influences the electromagnetic response of a type-II superconducting film to the applied magnetic field. We present a two-dimensional integrodifferential model for the magnetization of a flat superconductor/ferromagnet bilayer of an arbitrary shape using a thin shell quasistatic model for the ferromagnetic substrate and an infinitely thin approximation for the superconducting layer. An efficient numerical method is developed and used to investigate the effect of a ferromagnetic substrate. In particular, we simulate the thin bilayer magnetization in a parallel field and, for a high field, the critical-state distributions of the superconducting current density. These critical-state distributions are different from those known for a normal external field. Finally, we investigate the influence of a ferromagnetic substrate on superconducting film-based magnetic traps for cold atoms.**

## I. INTRODUCTION

Coated conductors are thin multilayer tapes, usually 4-12 mm wide, containing, in particular, about 1 µm thick layer of a type-II superconducting material upon a 30-100 µm thick metallic substrate. In the form of pancake coils or stacks of flat films, cut out from such tapes, these conductors are successfully used, e.g., for field trapping in superconducting magnets and magnetic shielding; the medical applications include MRI machines requiring high magnetic field uniformity [1-4]. Both nonmagnetic (Hastelloy) and magnetic substrates (typically, a Ni-W alloy) are used for manufacturing coated conductors. Substrates made of a Ni-W alloy, a soft ferromagnetic material, have good mechanical properties and can be cheaper [5]. Ferromagnetic substrates have an impact on the electromagnetic response of a coated conductor to the applied field or transport current. They can decrease AC losses in a perpendicular external field [6] and also hinder the reduction of the critical current in a strong field [7]. The presence of a ferromagnetic substrate leads to new effects, such as significant film magnetization and AC loss in a parallel external field. This explains the attention to studies of coated conductors with ferromagnetic substrates (see, e.g., [6, 8-11] and the references therein).

Analytically, the magnetization problem for an infinitely long coated conductor with a magnetic substrate was solved by Mawatari [12], who assumed Bean's critical-state model [13] for the superconductor and infinite magnetic permeability for the substrate. Various model formulations were employed in [8-11] to numerically simulate the magnetization of long coated conductors and their stacks, using more general constitutive relations for both the superconducting and ferromagnetic layers.

Flux penetration into superconducting films of different shapes on a ferromagnetic substrate has not been investigated yet, although for films on a nonmagnetic substrate this has attracted much interest and was studied in detail both experimentally (see, e.g., [14]) and numerically (see [15-17] and the references therein).

Here, we present a new model formulation and an efficient numerical method for simulating the magnetization of flat high-temperature superconducting films of an arbitrary shape upon ferromagnetic substrates.

For simplicity, we assume the ferromagnetic material has a constant magnetic permeability and neglect the eddy current in the substrate. We use the infinitely thin film representation and a nonlinear current-voltage relation for the superconducting layer, as is usual in the macroscopic superconducting film simulations [12, 15-18]. Although the ferromagnetic layer is much thicker than the superconducting layer, its thickness $\delta$ is typically much smaller than its other dimensions. The ferromagnetic layer can, therefore, be modeled using Krasnov's theory of quasistatic thin shell and plate magnetization [19-21]; schematically, this is illustrated in fig. 1.

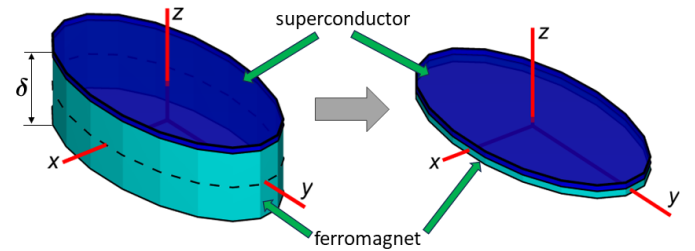

Fig. 1. Transition from three-dimensional to two-dimensional thin shell model of the ferromagnetic substrate.

Using this approach and the nonlinear eddy current model for the superconductor we obtain an evolutionary integrodifferential system of equations written for the superconducting sheet current density and the so-called "surface magnetization" of the substrate. Both variables are defined on the sample mid-surface, hence it is a two-dimensional (2D) problem. The advantage of this formulation is that no outer space consideration is necessary; this simplifies the numerical solution. A similar approach for modeling infinitely long strips was recently proposed in [22].

The arising one-dimensional problem was there solved by a very accurate spectral method which, however, cannot be extended to 2D problems considered here.

We express the sheet current density via its stream function and approximate the latter using continuous piecewise linear finite elements. Raviart-Thomas elements of the lowest order are employed for the surface magnetization. The method of lines is then applied for integration over time, and this allows us to account for a highly nonlinear current-voltage relation, characterizing a superconductor, without iterations. Numerical simulations show that the method is efficient; we use it to model cold atom traps on superconducting chips with a ferromagnetic substrate.

A limiting sheet current density distribution is reached asymptotically in a growing normal external field. This critical state exemplifies the key properties of thin film current density distributions [14, 15]. For a nonmagnetic substrate, the limiting critical-state solution can be found analytically; it is expressed via the distance-to-boundary function [16]. We found that a ferromagnetic substrate inhibits flux penetration into the superconducting film but does not alter the limiting solution. The situation is different for a growing field parallel to the film. If the substrate is nonmagnetic, the influence of a parallel field can usually be neglected. However, if the substrate is ferromagnetic, there establishes a critical sheet current density distribution, now expressed via the Kantorovich potential of the Monge-Kantorovich optimal transportation problem.

## II. SUPERCONDUCTING FILM ON A FERROMAGNETIC SUBSTRATE

Let the substrate occupy the domain $S \times (-\delta/2, \delta/2) \subset R^3$, where $\delta$ is the substrate thickness and $S$ is an open bounded 2D domain of the characteristic size $a \gg \delta$. We assume the substrate is an ideal soft magnet, nonconducting and having a constant relative magnetic permeability $\mu_r \gg 1$; below, we will use the susceptibility $\chi = \mu_r - 1$. The superconducting layer is deposited on one of the substrate surfaces, $z = -\delta/2$ or $z = \delta/2$, and is assumed infinitely thin. This layer is characterized by a nonlinear current-voltage relation; in our simulations, we will use the popular power law,

$$e = e_0 |j/j_c|^{n-1} j/j_c, \qquad (1)$$

where $e$ is the electric field, $j$ is the sheet current density, the power $n$ and the critical sheet current density $j_c$ are assumed constant, and the critical electric field $e_0 = 10^{-4}$ V/m.

The total magnetic field in this problem is the superposition, $h = h^e + h^{fm} + h^{sc}$, where $h^e$ is the given external field, $h^{fm}$ is the field induced by substrate magnetization, and $h^{sc}$ is the supercurrent-induced field.

To describe substrate magnetization, we use the thin shell model [19-21] obtained in the limit $\delta \to 0$, $\chi \to \infty$ while the product $\chi\delta$ remains constant. The model is justified if $a \gg \delta$ and $\chi \gg 1$. It is written for the surface magnetization function $\sigma$ defined on $S \times \{0\}$ as

$$\sigma = \int_{-\delta/2}^{\delta/2} m_\tau \mathrm{d}z,$$

where $m_\tau = (m_x, m_y)$ is the parallel to mid-surface component of the magnetization $m = \chi h$. The field $h^{fm}$ in this model is expressed as

$$h^{fm}(t,s) = \nabla \int_S \frac{\nabla' \cdot \sigma(t,s') \mathrm{d}s'}{4\pi r_{s's}}, \qquad (2)$$

where $r_{s's}$ is the distance between the mid-surface point $s' \in S \times \{0\}$ and a point $s \in R^3$. For a given field $h^0$ applied to the ferromagnet layer, the surface magnetization is determined by the integrodifferential equation

$$(\chi\delta)^{-1} \sigma(t,s) - \nabla \int_S \frac{\nabla' \cdot \sigma(t,s') \mathrm{d}s'}{4\pi r_{s's}} = h_\tau^0(t,s) \qquad (3)$$

for $s \in S \times \{0\}$ with the boundary condition

$$\sigma \cdot n = 0 \qquad (4)$$

on the boundary $\partial S$. In our case $h_\tau^0$ is the tangential component of $h^e + h^{sc}$ at the mid-surface. Taking into account the known properties of a single layer potential, we find that the tangential component of $h^{fm}$ is continuous, whereas its normal component takes different values on the two mid-surface sides:

$$h_z^{fm}\Big|_{z=0+} = -\frac{\nabla \cdot \sigma}{2}, \quad h_z^{fm}\Big|_{z=0-} = \frac{\nabla \cdot \sigma}{2}. \qquad (5)$$

By the Bio-Savart law, the field induced by the superconducting current is

$$h^{sc}(t,s) = \nabla \times \int_S \frac{j(t,s') \mathrm{d}s'}{4\pi r_{s's}}. \qquad (6)$$

This field has a continuous normal component; its tangential component values on the superconducting layer sides are different:

$$h_\tau^{sc}\Big|_{z=0+} = -\frac{\mathbf{1}_z \times j}{2}, \quad h_\tau^{sc}\Big|_{z=0-} = \frac{\mathbf{1}_z \times j}{2}, \qquad (7)$$

where $\mathbf{1}_z$ is the unit vector in the $z$-axis direction.

In (5) and (7) we neglected the substrate thickness, since $\delta \to 0$. However, the placement of the superconducting layer should be taken into account. Let us set $\nu = -1$ if the superconducting layer is above the substrate and $\nu = 1$ if it is below. Then, according to (5), in the superconducting layer we have $h_z = h_z^e + h_z^{sc} + (\nu/2)\nabla \cdot \sigma$. Substituting (6) and using the Faraday law $\mu_0 \partial_t h = -\nabla \times e$, for this layer we obtain

$$\mu_0 \partial_t \left( h_z^e + \nabla \times \int_S \frac{\boldsymbol{j}(t,s')\mathrm{d}s'}{4\pi r_{s's}} + \frac{v \nabla \cdot \boldsymbol{\sigma}}{2} \right) = -\nabla \times \boldsymbol{e}. \quad (8)$$

Here $\mu_0$ is the magnetic permeability of vacuum, $s \in S \times \{0\}$, and $\nabla \times \boldsymbol{f} := \partial_x f_y - \partial_y f_x$ is the scalar 2D curl operator.

Taking (7) into account, we rewrite (3) as

$$(\chi\delta)^{-1}\boldsymbol{\sigma}(t,s) - \nabla \int_S \frac{\nabla' \cdot \boldsymbol{\sigma}(t,s')\mathrm{d}s'}{4\pi r_{s's}} = \boldsymbol{h}_\tau^e - v\frac{\boldsymbol{1}_z \times \boldsymbol{j}(t,s)}{2}. \quad (9)$$

In this equation $s$ belongs to the mid-surface and the operator $\nabla$ is 2D.

If no transport current is applied, the sheet current density must meet the conditions
$$\nabla \cdot \boldsymbol{j} = 0 \text{ in } S, \quad \boldsymbol{j} \cdot \boldsymbol{n} = 0 \text{ on } \partial S. \quad (10)$$

Given an initial condition for $\boldsymbol{j}$, equations (1), (4), and (8)-(10) describe how magnetic flux penetrates into a superconducting layer on a ferromagnetic substrate. Note that for $v=0$ the system describes independent magnetization of each of the two layers.

To solve this system numerically, we reformulate it using a change of variables. To begin, let us proceed to the dimensionless variables

$$\tilde{\boldsymbol{j}} = \frac{\boldsymbol{j}}{j_c}, \quad \tilde{\boldsymbol{\sigma}} = \frac{\boldsymbol{\sigma}}{aj_c}, \quad \tilde{\boldsymbol{h}} = \frac{\boldsymbol{h}}{j_c},$$
$$\tilde{\boldsymbol{e}} = \frac{\boldsymbol{e}}{e_0}, \quad (\tilde{x},\tilde{y}) = \frac{(x,y)}{a}, \quad \tilde{t} = \frac{t}{t_0}, \quad (11)$$

where $t_0 = a\mu_0 j_c / e_0$. Omitting the sign "~" to simplify the notations, we rewrite (1), (8), (9) in dimensionless form as

$$\boldsymbol{e} = |\boldsymbol{j}|^{n-1}\boldsymbol{j}, \quad (12)$$

$$\partial_t \left( h_z^e + \nabla \times \int_S \frac{\boldsymbol{j}(t,s')\mathrm{d}s'}{4\pi r_{s's}} + \frac{v\nabla \cdot \boldsymbol{\sigma}}{2} \right) = -\nabla \times \boldsymbol{e}, \quad (13)$$

$$\kappa^{-1}\boldsymbol{\sigma} - \nabla \int_S \frac{\nabla' \cdot \boldsymbol{\sigma}(t,s')}{4\pi r_{s's}}\mathrm{d}s' = \boldsymbol{h}_\tau^e - v\frac{\boldsymbol{1}_z \times \boldsymbol{j}}{2}, \quad (14)$$

where $\kappa = \chi\delta/a$ is the dimensionless parameter, solely characterizing the substrate in this model. The form of equations (4) and (10) remains unchanged.

The magnetic susceptibility $\chi$ is zero for non-magnetic materials; for ferromagnetic substrates it is, usually, of the order of tens but can be much higher (thousands and even more) [23]. For, e.g., the strip width $2a = 10$ mm and substrate thickness $\delta = 100$ μm we have $\delta/a = 0.02$. Obviously, there is a wide range of possible $\kappa$ values. It was observed for long coated conductors in [22] that with the growth of $\kappa$ the thin shell model solutions tend to a limiting one and are practically the same for, e.g., all $\kappa > 10$.

For simplicity, we will now assume the domain $S$ is simply connected, although the formulation below can be extended to domains with holes (see, e.g., [16, 17]). By (10), there exists a stream function $g$ such that $g|_{\partial S} = 0$ and $\overline{\nabla} \times g = \boldsymbol{j}$ in $S$; here $\overline{\nabla} \times g := (\partial_y g, -\partial_x g)$ is the vectorial 2D curl. Below, we will plot current contours as the levels of $g$.

It is also convenient to use the rotated electric field $\boldsymbol{u} = (e_y, -e_x)$. In these new variables, equation (12) takes the form

$$\boldsymbol{u} = -|\nabla g|^{n-1}\nabla g. \quad (15)$$

Finally, since $\nabla \times \boldsymbol{e} = \nabla \cdot \boldsymbol{u}$ and $\boldsymbol{1}_z \times \boldsymbol{j} = \nabla g$, we rewrite equations (13) and (14) with the corresponding boundary conditions as

$$\begin{cases} \partial_t \left( h_z^e + \nabla \times \int_S \frac{\overline{\nabla} \times g(t,s')\mathrm{d}s'}{4\pi r_{s's}} + \frac{v\nabla \cdot \boldsymbol{\sigma}}{2} \right) = -\nabla \cdot \boldsymbol{u}, \\ g|_{\partial S} = 0, \end{cases} \quad (16)$$

$$\begin{cases} \kappa^{-1}\boldsymbol{\sigma} - \nabla \int_S \frac{\nabla' \cdot \boldsymbol{\sigma}(t,s')}{4\pi r_{s's}}\mathrm{d}s' = \boldsymbol{h}_\tau^e - \frac{v\nabla g}{2}, \\ \boldsymbol{\sigma} \cdot \boldsymbol{n}|_{\partial S} = 0. \end{cases} \quad (17)$$

Magneto-optic imaging is a useful technique for observing the magnetic flux distribution in superconducting films. It is commonly employed to examine the impact of film shape, defects, and inhomogeneities [21]. To compare simulation results with magnetic imaging observations, it is necessary to calculate the normal magnetic field component on the surface of the superconducting layer. In our model this component is

$$h_z = h_z^e + \nabla \times \int_S \frac{\boldsymbol{j}(t,s')\mathrm{d}s'}{4\pi r_{s's}} + \frac{v\nabla \cdot \boldsymbol{\sigma}}{2}. \quad (18)$$

### III. NUMERICAL METHOD

Let $\hat{S}$ be a polygonal approximation of $S$, $\mathcal{T} = \{\tau_i, i = 1,...,N_\tau\}$ - its regular triangulation with triangles $\tau_i$, $\mathcal{P} = \{p_i, i = 1,...,N_p\}$ and $\mathcal{E} = \{l_i, i = 1,...,N_e\}$ be, respectively, the sets of inner (not belonging to the boundary $\partial \hat{S}$) nodes and edges of this triangulation. We seek the approximate solution in the form

$$g = \sum_{i=1}^{N_p} g_i(t)\varphi_i(s), \quad \boldsymbol{\sigma} = \sum_{i=1}^{N_e} \sigma_i(t)\boldsymbol{\psi}_i(s). \quad (19)$$

Here $s \in \hat{S}$, $\varphi_i$ is the continuous scalar function linear in each triangular element, equal to one in the node $p_i \in \mathcal{P}$ and zero in all other nodes, and the vectorial function $\boldsymbol{\psi}_i$ is the lowest order Raviart-Thomas function [24] associated with the edge $l_i \in \mathcal{E}$ dividing two triangles, $\tau_i^\pm \in \mathcal{T}$. Let $|\tau_i|$ denote the area of $\tau_i$, $|l_i|$ - the length of $l_i$, and $p_i^{\pm}$ be the opposite to $l_i$ vertex in $\tau_i^{\pm}$. Then, denoting by $s$ the radius-vector of the point $s$, we can write

$$\boldsymbol{\psi}_i(s) = \begin{cases} \pm\dfrac{|l_i|}{2|\tau_i^{\pm}|}(s - p_i^{\pm}) & s \in \tau_i^{\pm}, \\ 0 & s \notin \tau_i^{\pm}, \end{cases}$$

and

$$\nabla \cdot \boldsymbol{\psi}_i = \begin{cases} \pm\dfrac{|l_i|}{|\tau_i^{\pm}|} & s \in \tau_i^{\pm}, \\ 0 & s \notin \tau_i^{\pm}. \end{cases}$$

These functions, also known as Rao-Wilton-Glisson basis functions, satisfy $\boldsymbol{\psi}_i|_{l_i} \cdot \boldsymbol{n}_i = 1$, where $\boldsymbol{n}_i$ is the unit normal to $l_i$ directed outside $\tau_i^+$, and $\boldsymbol{\psi}_i|_{l_k} \cdot \boldsymbol{n}_k = 0$ if $\boldsymbol{n}_k$ is a normal to any other edge $l_k \in \mathcal{E}$ or a boundary edge.

By using the basis functions $\varphi_j, \boldsymbol{\psi}_k$ associated with, respectively, only the inner nodes and inner edges, we satisfy the zero boundary conditions in (16) and (17). The same functions are used as the test functions. Multiplying (16) and (17) by the test functions, integrating, using Green's theorem, and noting that $(\overline{\nabla} \times g, \overline{\nabla} \times \varphi) = (\nabla g, \nabla \varphi)$, we obtain $N_p + N_e$ equations:

$$\iint_{\hat{S}\hat{S}} \frac{\nabla \dot{g}(t,s') \cdot \nabla \varphi_i(s)}{4\pi |s-s'|} ds'ds + \frac{v}{2}(\nabla \cdot \dot{\boldsymbol{\sigma}}, \varphi_i) = \qquad (20)$$
$$(\boldsymbol{u}, \nabla \varphi_i) - (\dot{h}_z^e, \varphi_i), \quad i = 1,...,N_p$$

$$\kappa^{-1}(\boldsymbol{\sigma}, \boldsymbol{\psi}_i) + \iint_{\hat{S}\hat{S}} \frac{[\nabla' \cdot \boldsymbol{\sigma}(t,s')][\nabla \cdot \boldsymbol{\psi}_i(s)]}{4\pi |s-s'|} ds'ds + \qquad (21)$$
$$\frac{v}{2}(\nabla g, \boldsymbol{\psi}_i) = (\boldsymbol{h}_\tau^e, \boldsymbol{\psi}_i), \quad i = 1,...,N_e.$$

Here $(.,.)$ denotes the scalar product of two functions, $\dot{f}$ means $\partial_t f$, and $\boldsymbol{u}$ is determined by (15).

Using the method of lines, we need to discretize only in space, which leads to a system of ordinary differential equations (ODE) with respect to time. For reader's convenience, we briefly describe the assembling of finite element matrices.

To handle the singular double-surface integrals in (20) and (21), we calculate first the symmetric $N_\tau \times N_\tau$ matrix with elements

$$K_{ij} = \iint_{\tau_i \tau_j} \frac{dsds'}{4\pi |s-s'|}. \qquad (22)$$

Here we follow the approach in [25]. In the most singular case of coinciding triangles, the elements $K_{ii}$ are computed exactly using the analytical expression in [26]. For touching or close elements, the analytical expression [27] is used for the inner integral (over $\tau_j$). The resulting nonsingular function is then integrated numerically over $\tau_i$ using the seven-point Gauss formula for triangles. Finally, for distantly separated elements, the integrals are approximated as $|\tau_i||\tau_j|/(4\pi r_{ij})$, where $r_{ij}$ is the distance between the element centers.

Let $\overline{g} = (g_1,...,g_{N_p})^T$ and $\overline{\boldsymbol{\sigma}} = (\sigma_1,...,\sigma_{N_e})^T$ denote the vectors of coefficients in (19). The $N_e \times N_e$ matrix with the elements

$$K_{ij}^\psi = \iint_{\hat{S}\hat{S}} \frac{\nabla \cdot \boldsymbol{\psi}_i(s) \nabla \cdot \boldsymbol{\psi}_j(s')}{4\pi |s-s'|} dsds'$$

is calculated as $K^\psi = CKC'$, where $C$ is the $N_e \times N_\tau$ matrix (see [24]) such that $C'\overline{\boldsymbol{\sigma}} = \nabla \cdot \boldsymbol{\sigma}$, and the piece-wise constant $\nabla \cdot \boldsymbol{\sigma}$ is regarded as a vector. Similarly, presenting the derivatives $\partial_x g, \partial_y g$, constant in each triangle, as $N_\tau \times 1$ vectors and denoting by $G_x, G_y$ such $N_\tau \times N_p$ matrices that $G_x \overline{g} = \partial_x g, G_y \overline{g} = \partial_y g$, we compute the $N_p \times N_p$ matrix $K^\varphi = G_x' K G_x + G_y' K G_y$ with the elements

$$K_{ij}^\varphi = \iint_{\hat{S}\hat{S}} \frac{\nabla \varphi_i(s) \cdot \nabla \varphi_j(s')}{4\pi |s-s'|} dsds'.$$

The products $(\boldsymbol{\sigma}, \boldsymbol{\psi}_i)$ in (21) are elements of the vector $B\overline{\boldsymbol{\sigma}}$, where $B$ is the $N_e \times N_e$ matrix with $B_{ij} = (\boldsymbol{\psi}_i, \boldsymbol{\psi}_j)$, see [24]. Let $D_x, D_y$ be $N_\tau \times N_e$ matrices such that the $i$-th row of $[D_x\overline{\boldsymbol{\sigma}}, D_y\overline{\boldsymbol{\sigma}}]$ gives the value of $\boldsymbol{\sigma}$ at the center of $\tau_i$. Then the terms $(\nabla g, \boldsymbol{\psi}_i)$ in (21) can be presented as $[D_x' D_\tau G_x + D_y' D_\tau G_y]\overline{g}$, where $D_\tau$ is the diagonal matrix with $D_{\tau,jj} = |\tau_j|$. The term $(\nabla \cdot \dot{\boldsymbol{\sigma}}, \varphi_i)$ in (20) is, in the matrix form, represented by $FC'\dot{\overline{\boldsymbol{\sigma}}}$, where $F$ is the $N_p \times N_\tau$ matrix with elements $F_{ij} = |\tau_j|/3$ if the node $p_i$ is a vertex of $\tau_j$ and zero otherwise.

Finally, for the uniform external fields $\boldsymbol{h}^e = \boldsymbol{h}^e(t)$ in our simulations, the right-hand sides of (20) and (21), respectively, are the vectors

$$\overline{R}^\varphi = G_x' D_\tau \overline{u}_x + G_y' D_\tau \overline{u}_y - \dot{h}_z \overline{S}/3,$$
$$\overline{R}^\psi = h_x^e D_x' \overline{\tau} + h_y^e D_y' \overline{\tau}. \qquad (23)$$

Here the vectors $\overline{u}_x = \overline{u}_x(\overline{g}), \overline{u}_y = \overline{u}_y(\overline{g})$ contain, respectively, the values of $|\nabla g|^{n-1} \partial_x g$, $|\nabla g|^{n-1} \partial_y g$ in the triangles, $S_i$ is the sum of areas of all triangles having $p_i$ as a vertex, and $\overline{\tau} = (|\tau_1|,...,|\tau_{N_\tau}|)^T$.

Equation (21) can now be written in the matrix form as $A_{11}\bar{\sigma}+A_{12}\bar{g}=\bar{R}^{\psi}$, where $A_{11}=\kappa^{-1}B+K^{\psi}$ and $A_{12}=(\nu/2)\left[D'_{x}D_{\tau}G_{x}+D'_{y}D_{\tau}G_{y}\right]$. Inverting $A_{11}$, we substitute

$$\bar{\sigma}=A_{11}^{-1}\left(\bar{R}^{\psi}-A_{12}\bar{g}\right) \quad (24)$$

into (20), written as $A_{21}\dot{\bar{\sigma}}+A_{22}\dot{\bar{g}}=\bar{R}^{\varphi}$ with $A_{21}=(\nu/2)FC'$, $A_{22}=K^{\varphi}$. Thus, we obtain

$$\dot{\bar{g}}=(A_{22}-A_{21}A_{11}^{-1}A_{12})^{-1}\left[\bar{R}^{\varphi}-A_{21}A_{11}^{-1}\dot{\bar{R}}^{\psi}\right].$$

Denoting $A=(A_{22}-A_{21}A_{11}^{-1}A_{12})^{-1}$ and substituting expressions (23), we arrive at the ODE system written for the stream function $g$ alone:

$$\dot{\bar{g}}=U_{x}\bar{u}_{x}(\bar{g})+U_{y}\bar{u}_{y}(\bar{g})+\dot{h}^{e}_{z}\bar{V}_{z}+\dot{h}^{e}_{x}\bar{V}_{x}+\dot{h}^{e}_{y}\bar{V}_{y}, \quad (25)$$

where

$$U_{x}=AG'_{x}D_{\tau}, \quad U_{y}=AG'_{y}D_{\tau}, \quad \bar{V}_{z}=-A\bar{S}/3,$$
$$\bar{V}_{x}=-AA_{21}A_{11}^{-1}D'_{x}\bar{\tau}, \quad \bar{V}_{y}=-AA_{21}A_{11}^{-1}D'_{y}\bar{\tau},$$

and, for a known $\bar{g}(t)$, the vectors $\bar{u}_{x},\bar{u}_{y}$, are determined by (15) with $\partial_{x}g=G_{x}\bar{g}$ and $\partial_{y}g=G_{y}\bar{g}$.

Provided the solution $\bar{g}$ of (25) is found, the approximate sheet current density components are computed as $\bar{j}_{x}=G_{y}\bar{g}, \bar{j}_{y}=-G_{x}\bar{g}$, and the surface magnetization is determined by (24).

The magnetic field can be also found. In particular, to compute the normal magnetic field component (18), we multiplied this equation by the test functions and integrated, which yielded

$$(h_{z},\varphi_{i})=(h^{e}_{z},\varphi_{i})+$$
$$\iint_{\bar{s}\;\bar{s}}\frac{\nabla g(t,s')\cdot\nabla\varphi_{i}(s)}{4\pi|s-s'|}\,\mathrm{d}s'\mathrm{d}s+\frac{\nu}{2}(\nabla\cdot\boldsymbol{\sigma},\varphi_{i}).$$

Discretizing, we find this field in the inner mesh nodes: $\bar{h}_{z}=h^{e}_{z}\bar{I}+3\mathrm{diag}(\bar{S})^{-1}\left[K^{\varphi}\bar{g}+A_{21}\bar{\sigma}\right]$, where the $N_{p}\times 1$ vector $\bar{I}=(1,...,1)^{T}$. We note that in the infinitely thin film approximation employed here, the field $h_{z}$ on the film boundary is, typically, infinite (has a logarithmic singularity).

To calculate the field $\boldsymbol{h}=\boldsymbol{h}^{e}+\boldsymbol{h}^{sc}+\boldsymbol{h}^{fm}$ in the space outside the film, we rewrite (2) and (6) as, respectively,

$$\boldsymbol{h}^{fm}=-\frac{1}{4\pi}\int_{S}\nabla'\cdot\boldsymbol{\sigma}(t,\boldsymbol{r}')\frac{\boldsymbol{s}-\boldsymbol{s}'}{|\boldsymbol{s}-\boldsymbol{s}'|^{3}}\mathrm{d}s',$$

$$\boldsymbol{h}^{sc}=\frac{1}{4\pi}\int_{S}\boldsymbol{j}(t,s')\times\frac{\boldsymbol{s}-\boldsymbol{s}'}{|\boldsymbol{s}-\boldsymbol{s}'|^{3}}\mathrm{d}s',$$

substitute $\boldsymbol{j}=\left(\partial_{y}g,-\partial_{x}g\right)$, and, after discretization, obtain

$$\boldsymbol{h}(t,s)=\boldsymbol{h}^{e}+$$
$$\sum_{i=1}^{N_{\tau}}\left\{\begin{array}{c}-\alpha_{i}(s)\left[\partial_{x}g\big|_{e_{i}}\right]-\beta_{i}(s)\left[\nabla\cdot\boldsymbol{\sigma}\big|_{e_{i}}\right]\\ -\alpha_{i}(s)\left[\partial_{y}g\big|_{e_{i}}\right]-\gamma_{i}(s)\left[\nabla\cdot\boldsymbol{\sigma}\big|_{e_{i}}\right]\\ \beta_{i}(s)\left[\partial_{x}g\big|_{e_{i}}\right]+\gamma_{i}(s)\left[\partial_{y}g\big|_{e_{i}}\right]-\alpha_{i}(s)\left[\nabla\cdot\boldsymbol{\sigma}\big|_{e_{i}}\right]\end{array}\right\}, \quad (26)$$

where

$$\alpha_{i}(s)=\frac{z}{4\pi}\int_{\tau_{i}}\frac{\mathrm{d}s'}{|\boldsymbol{s}-\boldsymbol{s}'|^{3}}, \quad \beta_{i}(s)=\frac{1}{4\pi}\int_{\tau_{i}}\frac{(x-x')\mathrm{d}s'}{|\boldsymbol{s}-\boldsymbol{s}'|^{3}},$$

$$\gamma_{i}(s)=\frac{1}{4\pi}\int_{\tau_{i}}\frac{(y-y')\mathrm{d}s'}{|\boldsymbol{s}-\boldsymbol{s}'|^{3}}.$$

For an outside-the-film point $s$ these integrals are non-singular and coefficients $\alpha_{i},\beta_{i},\gamma_{i}$ can be calculated numerically. Even a one-point quadrature should be sufficient if the point $s=(x,y,z)$ is not too close to the film. For the close-to-film points it can be better to find these integrals analytically (see [28] and the references therein).

## IV. SIMULATION RESULTS

Our numerical simulations were performed in Matlab R2020b on a PC with the Intel® Core™ i7-9700 CPU 3.00 GHz and 64GB RAM. The Matlab PDE toolbox was used for the finite element mesh generation; the ODE system (25) was solved using the Matlab ode23 solver with the relative and absolute tolerances both set to $10^{-6}$. We have assumed zero initial current $(g|_{t=0}=0)$ in all of our examples. The simulation results are presented in the dimensionless form (11).

To verify the convergence of numerical solution with the finite element mesh refinement, we solved the problem (25) assuming $S=\{|x|<1,|y|<1\}$, $\kappa=1, \nu=-1$, the normal external field $h^{e}_{z}=t$ and $n=100$, using two meshes. After interpolating the node values of $g$ into the points of a regular $100\times 100$ grid in $S$, we compared the obtained numerical solutions at $t=0.5$ (table 1). The solution, computed for a cruder mesh in one minute, is very close to that obtained using a much finer mesh. Even for such extremely nonlinear current-voltage relation the employed ODE solver was sufficiently efficient. Such a behavior was observed also in our examples below.

**Table 1.** Convergence of the numerical method.

| Maximal element size | $N_{\tau}$ | CPU time (min) | Difference in the $L^{\infty}$-norm | |
|---|---|---|---|---|
| | | | absolute | relative (%) |
| 0.05 | 5,536 | 1 | 0.0059 | 1.3 |
| 0.025 | 22,222 | 27 | | |

As a validation of our method, we solved the problem for an elongated rectangle $S = \{|x|<3, |y|<1\}$ in the normal external field $\boldsymbol{h}^e = \{0, 0, t\}$. The finite element mesh, generated with the element size not exceeding 0.075, consisted of 7334 elements $\tau_i$.

Sufficiently far from the rectangle sides $x = \pm 3$ the solution is expected to be similar to that for an infinitely long strip. We compared the calculated current density values in the central strip part to the solution of a corresponding 1D problem for three values of parameter $\kappa$. For $\kappa = 0.01$ the influence of a ferromagnetic substrate is negligible. The infinite strip problem with a nonmagnetic substrate has a known analytical solution [18] for the Bean critical-state model with the multivalued current-voltage relation

$$\boldsymbol{e} \parallel \boldsymbol{j}, \quad |\boldsymbol{j}| \le 1, \quad |\boldsymbol{j}| < 1 \to \boldsymbol{e} = 0 \qquad (27)$$

in the dimensionless variables. This model can be regarded as the $n \to \infty$ limit of models with the power current-voltage relation, see [29]. We chose the large power, $n = 100$, to compare our 2D problem solution with this 1D analytical one. Similarly, for large powers $n$ and values of $\kappa$, solution of the magnetization problem becomes close to that for the Bean model and $\kappa = \infty$. We found that $\kappa = 10$ is sufficiently large, and compared our 2D problem solution in the same central area to the corresponding analytical 1D-problem solution [12]. No analytical solution is available for $0 < \kappa < \infty$ and, for $\kappa = 0.5$, we compared our solution for the same $n$ with the numerical 1D-problem solution [22].

In each case, the computed sheet current density was practically parallel to the $x$ axis in the central strip part. Let $\{x_i^0, y_i^0\}$ be the center of $\tau_i$. Plotting $j_x|_{\tau_i}$ versus $y_i^0$ for elements with $|x_i^0| < 0.3$, we obtained a very good agreement with the 1D solutions (Fig. 2).

In these examples, computing the integrals (22) and assembling the finite element matrices took about 0.9 min.; from 0.8 to 1.2 minute was needed to the ODE solver to solve the system (25) for $0 \le t \le 0.6$.

Magnetic flux (18) penetrates into a superconducting layer from its boundary and is zero in the subcritical current density areas, where $|\boldsymbol{j}| < 1$. Our results indicate that ferromagnetic substrates impede flux penetration in a normal external field, as shown in Fig. 2. This is clearly seen also in our second example, where the domain $S$ has an irregular shape (Fig. 3).

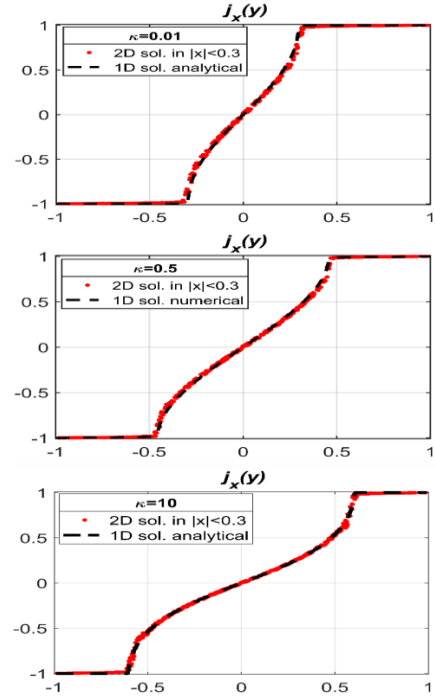

**Fig. 2.** Sheet current density in the central part of an elongated rectangle. Simulation results for the normal external field $h_z^e = 0.6$, $n=100$ and $\nu=1$. Numerical solutions of the 2D problem are compared with 1D problem solutions for an infinitely long strip: for $\kappa=0.01$ and $\kappa=10$ with the analytical solutions [18] and [12], respectively; for $\kappa=0.5$ - with the numerical solution [22].

The same normal field growing up to $h_z^e = 0.6$ is applied but now $n = 30$ (a more realistic value) and the mesh of 17434 elements is refined near the domain boundary to better represent the magnetic field $h_z$ in the superconducting layer (fig. 4). The field penetrates deeper into this layer on a nonmagnetic substrate (Fig. 3, bottom) but, qualitatively, solutions for different substrates are similar.

For a bilayer in a parallel external field the situation is different. Superconducting layers on nonmagnetic substrates remain almost unmagnetized. On the contrary, ferromagnetic substrates change the magnetic field direction and, as a result, external field variations can induce in the film a superconducting current.

We simulated the magnetization of a circular and a cross-shaped bilayers in the external field $\boldsymbol{h}^e = (0, t, 0)$ using the meshes, respectively, with 12 and 17.5 thousand finite elements (Fig. 5). In these examples $\kappa = 1, v = 1$ and the results are shown for $t = 2$. The simulations took 3.7 minutes for the circular domain and 12.6 minutes for the cross-like one.

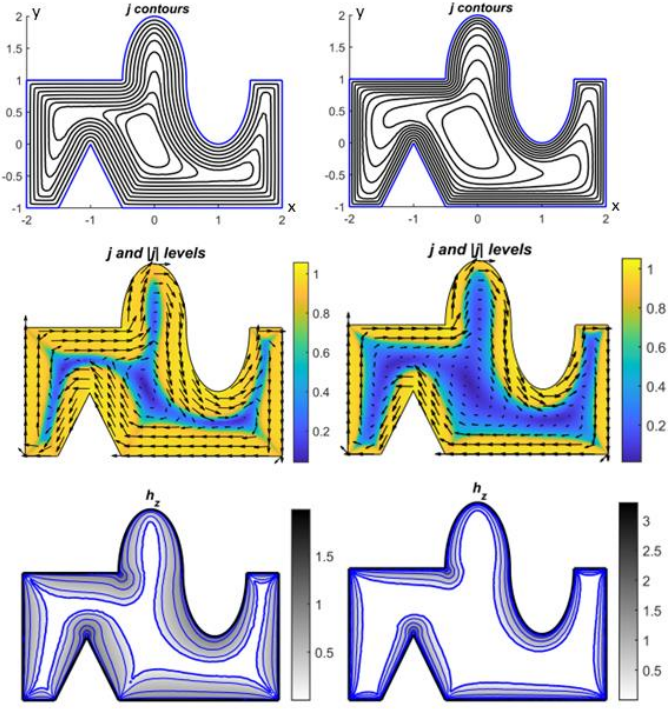

**Fig. 3.** Sheet current density (top and middle) and levels of the normal magnetic field (bottom) in an irregularly shaped film, $n = 30$. Left: $\kappa=0.01$, an almost nonmagnetic substrate. Right: $\kappa=10$, a strongly magnetic one.

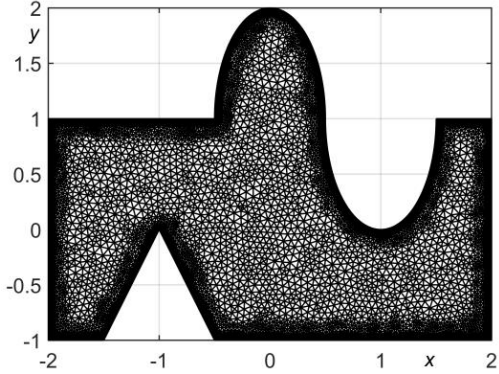

Fig. 4. The finite element mesh.

As our next example, we considered a square sample in the parallel field that grows as $\boldsymbol{h}^e = (0, 3t, 0)$ for $0 < t < 1$ and for $1 \leq t \leq 1.5$ rotates in the film plane as $\boldsymbol{h}^e = (3\sin(2\pi t), 3\cos(2\pi t), 0)$. The computed sheet current density is presented for $t = 1, 1.25,$ and $1.5$ (Fig. 6). Solution of this problem took about 10 minutes for the finite element mesh with eleven thousand elements.

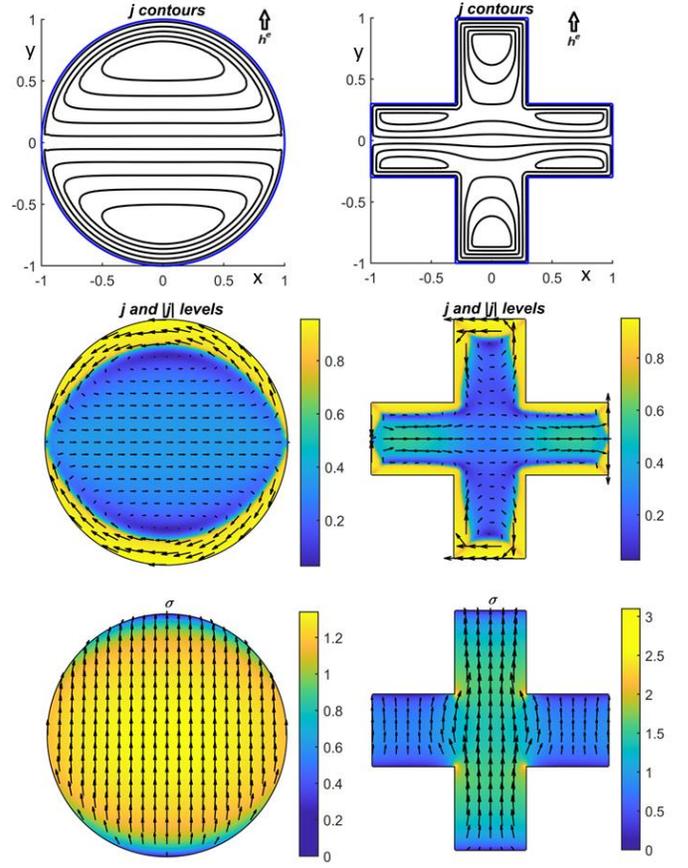

**Fig. 5.** Magnetization of two films in a parallel external field. Simulation results for $h_y^e = 2$, $n=30$, $\kappa = 1$, $v = 1$. Shown (from top to bottom): the current contours, sheet current densities, and substrate magnetizations.

Due to hysteretic superconductor magnetization, the solutions for the external field directions $0^\circ$ and $180^\circ$ are not symmetrical. The computations show that, after the field rotation by about $90^0$, there establishes a periodic solution rotating with the field.

It remains to investigate the influence of the layer placement parameter $v$. Let, as in all of our simulations, $\boldsymbol{j}|_{t=0} = 0$. Simple analysis of equations (12)-(14) shows that if $h_z^e = 0$ and the sign of $v$ changes, solution $(\boldsymbol{j}, \boldsymbol{\sigma})$ is replaced by $(-\boldsymbol{j}, \boldsymbol{\sigma})$. If $\boldsymbol{h}_\tau^e = 0$, the new solution is $(\boldsymbol{j}, -\boldsymbol{\sigma})$. Otherwise, if both the normal and tangential components of the external field are nonzero, the solution changes in a nontrivial way. As an example, we set $\boldsymbol{h}^e = (t, 0, t/2)$ and present the sheet current densities at $t = 1$ computed for $v = 1$ and $v = -1$ (Fig. 7). The surface magnetizations are also different in these two cases but their difference is less pronounced, since both of them are mainly controlled by $\boldsymbol{h}_\tau^e$.

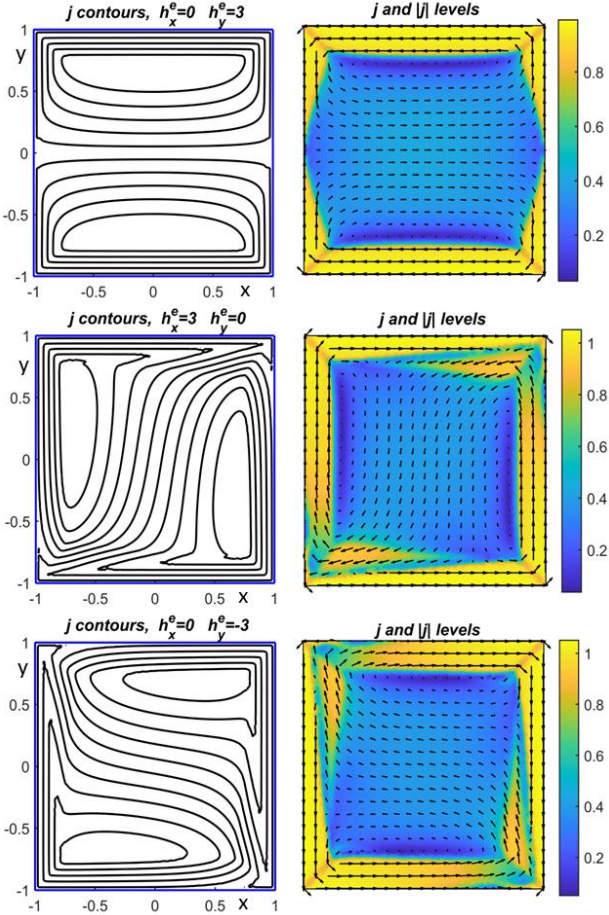

**Fig. 6.** Square sample in a rotating parallel external field. The current contours (left) and sheet current densities (right) are shown before the field rotation (top), after its rotation by 90° (middle), and by 180° (bottom). Here $n=30$, $\kappa=1$, $\nu=1$.

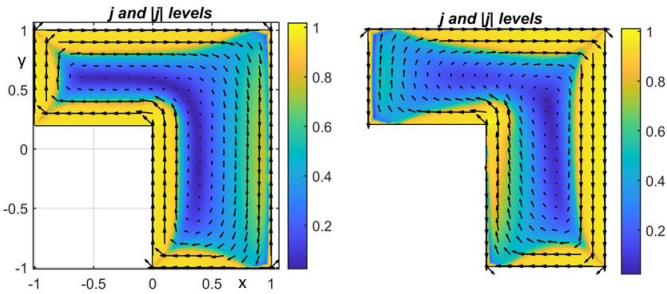

**Fig. 7.** Influence of the superconducting layer location. The sheet current density at $h^e=\{1,0,0.5\}$ is computed for $\nu=1$ (left) and $\nu=-1$ (right). Here $n=30$, $\kappa=1$.

## V. THE BEAN CRITICAL STATE

Let us consider the hybrid bilayer in a growing external field, e.g., $h^e(t,s) = t\hat{h}(s)$ and, for the superconducting layer, assume the Bean current-voltage relation (27). Since $j$ is bounded, there establishes a limiting sheet current density distribution: $j(t,s) \to j^\infty(s)$ as $t \to \infty$. Studying this limit aids, as in the case of a nonmagnetic substrate [14, 15], to understand the essential flux penetration properties.

With the growth of the external magnetic field, the impact of the current on the substrate magnetization decreases. Assuming $\sigma(t,s)/t \to \hat{\sigma}(s)$ as $t \to \infty$ and using equations (16)-(17), we obtain

$$\hat{h}_z + \frac{\nu \nabla \cdot \hat{\sigma}}{2} = -\nabla \cdot u^\infty, \qquad (28)$$

$$\kappa^{-1}\hat{\sigma} - \nabla \int_s \frac{\nabla' \cdot \hat{\sigma}(t,s')}{4\pi r_{s's}} ds' = \hat{h}_\tau, \quad \hat{\sigma} \cdot n\big|_{\partial S} = 0. \qquad (29)$$

Here (29) fully determines $\hat{\sigma}$, and $u^\infty = \{e_y^\infty, -e_x^\infty\}$ is the $t \to \infty$ limit of the rotated electric field. Employing the power law approximation of the Bean model and presenting $u$ via the stream function $g$, see (15), we rewrite (28) as a Dirichlet problem for the $p$–Laplacian equation (in usual notations, $p = n+1$),

$$-\nabla \cdot \left(|\nabla g_n|^{n-1} \nabla g_n\right) = f, \qquad g_n\big|_{\partial S} = 0, \qquad (30)$$

where $f = -\hat{h}_z - \nu\nabla \cdot \hat{\sigma}/2$. The function $g^\infty = \lim_{n\to\infty} g_n$ determines the limiting sheet current density distribution: $j^\infty = \overline{\nabla} \times g^\infty$. The $p \to \infty$ limit of the $p$–Laplacian equation is the problem arising in elasto-plasticity [30], optimal transportation [31], etc. It has been shown [32] that $g^\infty$ is a maximizer of $\int_S gf\,ds$ in the set

$$\left\{ g \in W^{1,\infty}(S): \quad g\big|_{\partial S} = 0, \quad \|\nabla g\|_{L^\infty(S)} \le 1 \right\}.$$

This problem coincides with the written for Kantorovich potential dual formulation of the unbalanced optimal transportation problem [32], in which the material from the pile $z = \max(f(s),0)$ should be removed and the pit $z = \min(f(s),0)$ filled in the cheapest way, and it is allowed to leave or pick up any needed amount of the material also at the domain boundary.

If $f > 0$, the well-known solution ([30], see also [29]) is the distance-to-boundary function $g^\infty(s) = \text{dist}(s,\partial S)$ (with the minus if $f < 0$) and, up to a set of measure zero, $|j^\infty| = |\nabla g^\infty| = 1$ (the critical state). Such is, e.g., the solution for a strong uniform normal external field: since $\hat{h}_\tau = 0$, also $\hat{\sigma} = 0$ [see (29)] and $f = -\hat{h}_z$. In this case the so-called $d^+$ lines [14, 15], at which the direction of $j^\infty$ changes discontinuously, are the ridges of $S$, consisting of points $s \in S$ for which the minimal distance to the boundary is reached in more than one boundary points. The same critical state establishes in a strong normal field in superconducting layers on nonmagnetic and ferromagnetic substrates.

If $\text{sign}(f)$ is not constant, $|j^\infty| = 1$ almost everywhere in

supp($f$), however, no general analytical solution is available. This is the case, e.g., if $\hat{h}_z = 0$, since then $f = -\nu \nabla \cdot \hat{\boldsymbol{\sigma}}/2$ and $\int_S f \, ds = 0$, so the sign of $f$ cannot be everywhere the same. To compute the critical-state solutions for films in a strong parallel field, we assumed $\boldsymbol{h}^e(t,s) = t\{\hat{h}_x(s), \hat{h}_y(s), 0\}$, chose the high power $n = 100$ to approximate the Bean model, and solved the evolutionary problem (25) until the time when the external field is sufficiently strong. The critical state does not depend on the value of $\kappa > 0$ but, if $\kappa$ is small, a stronger field is needed to reach a good approximation to the critical state. In our simulations we used a large value, $\kappa = 10$. Results of these simulations with $|\hat{\boldsymbol{h}}_\tau^e| = 10$ are presented for two differently shaped samples in Fig. 10 for $t = 1$. The $d^+$ lines (white) are clearly seen in the critical-state current density plots. The limiting critical-states in a parallel external field are completely different from the previously studied critical-states in a normal field.

## VI. SUPERCONDUCTING CHIP-BASED MAGNETIC TRAPS: THE INFLUENCE OF A FERROMAGNETIC SUBSTRATE

Ferromagnetic substrates of coated conductors can improve the characteristics of magnetic shields [1] and dynamo flux pumps [33]. Here we employ our numerical method to investigate the influence of a ferromagnetic substrate on superconducting chip-based magnetic traps for ultracold atoms. Such traps are regarded as excellent candidates for quantum information applications and quantum simulations (see, e.g. [34, 35] and the references therein).

Trapping of atoms having nonzero magnetic moments is based on the Zeeman effect: the atoms, whose potential energy increases with the magnetic field, can be confined and held near the local minimum of the field magnitude. To realize such a trap, the kinetic energy of atoms should be much lower than the depth of the trap potential well, and the gradient of the magnetic field should be sufficient to prevent gravity and other forces from pulling the atoms out of the trap. Modern technology allows one to reduce atom cloud temperature to tens of nano-Kelvins. The typical trapping criterion is $\tilde{\mu}\mu_0 h_{dep} \geq 10 k_B T_a$, where $\tilde{\mu}$ is the atom magnetic moment, $k_B$ is the Boltzmann constant, $T_a$ is the atom cloud temperature, and $h_{dep}$ is the trap depth, which is the difference between the maximal level of the magnetic field magnitude for which the iso-surface of $|\boldsymbol{h}|$ is closed, and the minimum of $|\boldsymbol{h}|$ inside the trap.

Various distributions of persistent superconducting currents can produce a necessary magnetic field configuration. Thus, two opposite pulses of the normal magnetic field were applied in [36] to induce in a $1 \times 1$ mm² superconducting film on a non-magnetic substrate a persistent current, for which the induced magnetic field is zero at some height above the square center. For the sequence of two $h_z^e$ pulses, approximately, in dimensionless variables, $0 \to 3 \to 0$ (the first pulse) and $0 \to -0.8 \to 0$ (the second pulse), a cloud of ultracold ⁸⁷Rb atoms was held in the created magnetic trap for about 1.5 s.

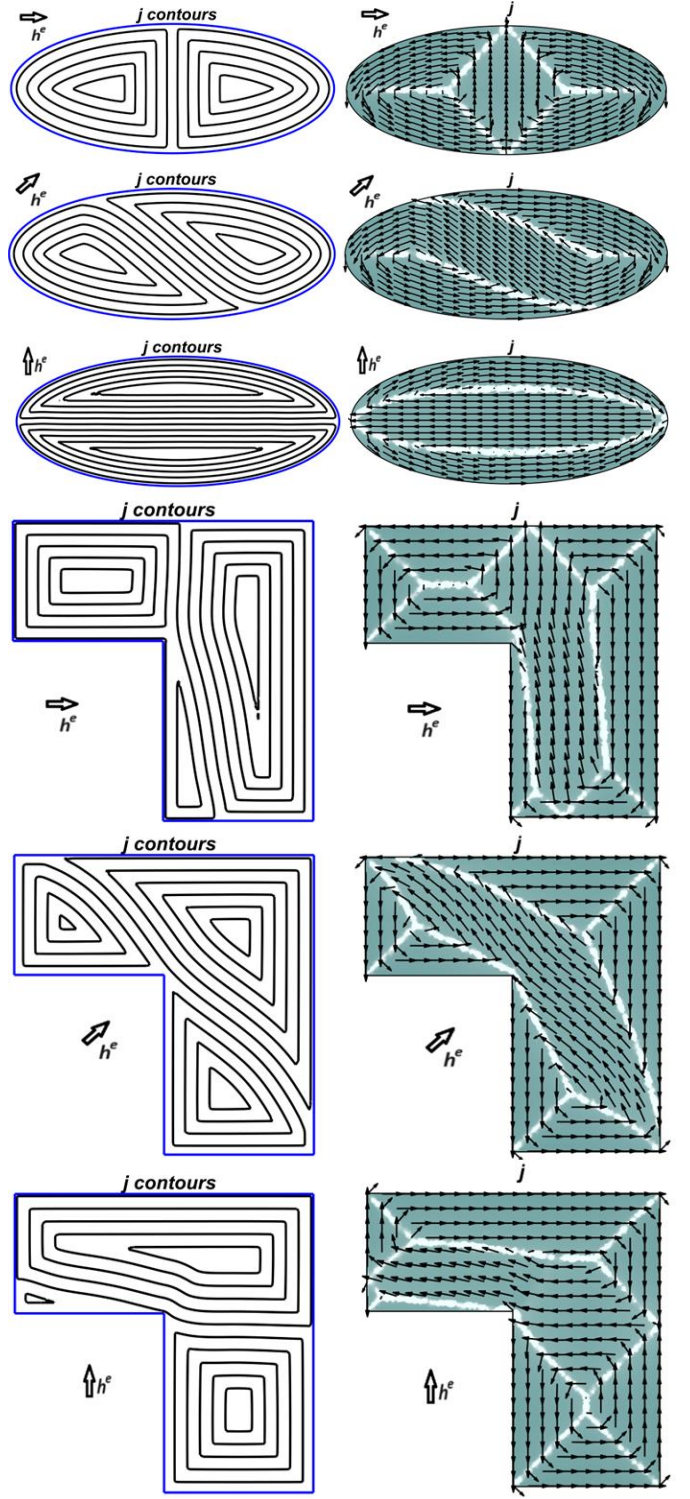

**Fig. 8.** Critical-state sheet current density. Simulation results for different direction of a strong parallel external field; $\nu = -1$.

Our numerical simulation [28] reproduced the main features of this trap (its height above the film, size, and shape) qualitatively well. The influence of a ferromagnetic substrate on the trap properties is not obvious *a priori* and here we investigate this numerically using the proposed magnetization model. Let a square $a \times a$ superconducting film be placed above the substrate. First, following [28], we chose a high power ($n = 100$) and assumed that a similar sequence of two pulses is applied. The dimensionless duration of this sequence in our simulations was $T \approx 7$. We note, however, that with such a high power the model is close to the rate-independent Bean model, so duration of pulses is not important. For $\kappa = 0.001$ the influence of the substrate is negligible and our simulation results are as in [28], where another numerical scheme (the mixed method with the implicit discretization in time) has been employed; the simulations showed that any significant change of the second pulse magnitude decreases the trap depth.

Now we simulated also traps with the substrate parameters $\kappa = 0.2$ and $\kappa = 0.4$. At the peak of the strong positive pulse, the films are in the same critical state with $|j| = 1$ almost everywhere (see the Section V). The penetration of current induced by the weaker negative pulse is hindered, as was shown above, by ferromagnetic substrates. To compensate this and obtain a current distribution similar to that in the case of a nonmagnetic substrate, we applied stronger negative pulses, $0 \to -1 \to 0$ for $\kappa = 0.2$ and $0 \to -1.2 \to 0$ for $\kappa = 0.4$. Our numerical simulations were performed with a mesh of 8500 finite elements and took about 45 minutes each. Computed persistent sheet current density after two pulses is shown in fig. 9 for $\kappa = 0.2$; for the two other values of $\kappa$ it is approximately the same.

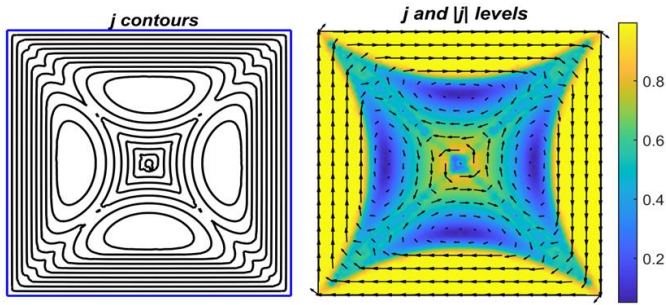

**Fig. 9.** Sheet current density (dimensionless) after two opposite pulses of the normal magnetic field; $\kappa = 0.2$, $v = -1$

The magnetic field magnitude and its closed iso-surfaces characterizing magnetic traps are shown in fig. 10. Results of this simulation suggest that deeper magnetic traps with steeper walls can be created using superconducting chips with ferromagnetic substrates.

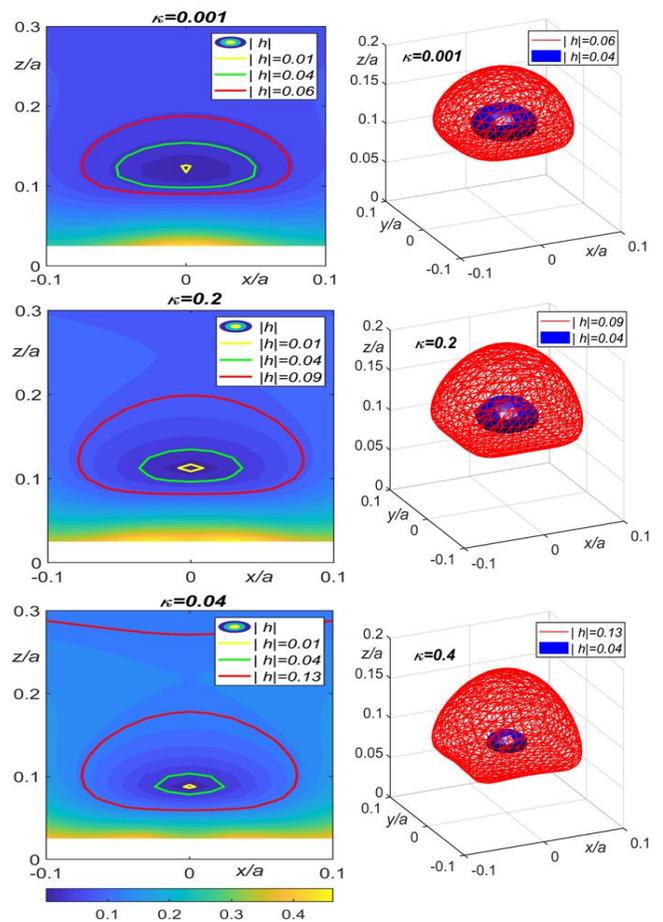

**Fig. 10.** Magnetic field magnitude $|h|$ above the film in the plane $y = 0$ (left) and closed $|h|$ iso-surfaces (right); dimensionless. Computed for a square superconducting film with three different substrates; $v = -1$. The maximal $|h|$ value, for which the iso-surface is closed (red), determines the trap depth.

## VII. Discussion

This paper extends the one-dimensional model for an infinitely long coated conductor with a ferromagnetic substrate, as described in reference [20], to flat samples of an arbitrary shape cut out from a coated conductor. Although the macroscopic magnetization of superconducting films on non-magnetic substrates is well studied, the influence of a ferromagnetic substrate has been much less investigated in the previous works. The main difficulties in numerical simulation of such hybrid systems, consisting of thin superconducting and ferromagnetic layers, are the system geometry, characterized by a very high aspect ratio, and strong nonlinearity of the constitutive relations.

The proposed model and its integrodifferential formulation employ the infinitely thin approximation for the superconducting layer coupled to the thin shell magnetization model for the ferromagnetic layer. The latter model has been developed in seminal but, probably, not well-known works by I.P. Krasnov [19, 20]. This approach resolves the high aspect ratio complication and makes the model two-dimensional.

Implicit discretization with respect to time is most often employed for numerical solution of the applied superconductivity problems. Sophisticated iterations are necessary then to deal with a highly nonlinear current-voltage relation characterizing the superconductor (see, e.g., [17, 37]). By utilizing the method of lines, we were able to delegate control over time integration accuracy to a standard ODE solver and avoid iterations. This approach is simpler and, as was demonstrated by numerical examples in this and other works [38, 39], can be efficient for different superconductivity problems.

The developed method was applied to simulation of a magnetic trap for cold atoms; we found that better traps can be created if the substrate is ferromagnetic.

To our knowledge, no similar 2D simulation has been done before. Our goal was to develop a model and numerical method allowing one to investigate the main effects related to a ferromagnetic substrate, like superconducting layer magnetization in a parallel field. Hence, we used the simplest constitutive relations allowing us to do this. Namely, we assumed the field-independent critical sheet current density $j_c$ and relative magnetic permeability $\mu_r$. Here we considered neither inhomogeneous nor multiply connected films. However, our numerical method can be extended to account for more complicated constitutive relations, film inhomogeneity, and the presence of holes. Our analysis was limited to the superconducting layer and the non-conducting ferromagnetic substrate. Usually, coated conductors include also tens micrometers thick normal metal (Ag, Cu) layers as electric and thermal stabilizers. Replacement of the current-voltage relation (1) by the relation that uses the effective resistivity of the superconductor, normal metal, and a conductive substrate connected in parallel can account for these layers. It is also possible to incorporate into the model an additional thin layer of normal conducting material and use the infinitely thin approximation for its description.

We note that on the mesoscopic level of magnetic vortices in the superconductor and magnetic domains in the ferromagnet, such hybrid bilayer was simulated in [40].